\documentclass[twocolumn,prl,preprintnumbers,amsmath,amssymb,
a4paper]{revtex4}
\usepackage[]{epsfig}
\newcommand{\dd}{\mathrm{d}}
\newcommand{\be}{\begin{equation}}
\newcommand{\ee}{\end{equation}}
\newcommand{\fav}{F_{\rm av}}
\newcommand{\s}{\sigma}
\newcommand{\dt}{\Delta t}

\newcommand{\tw}{t_\mathrm{w}}
\newcommand{\eq}{\mathrm{eq}}

\begin{document}
\title{Heterogeneous Dynamics of Coarsening Systems}
\author{P. Mayer,$^1$ H. Bissig,$^2$
L. Berthier,$^{3,4}$ L. Cipelletti,$^5$ J.P. Garrahan,$^{6}$ P.
Sollich,$^1$ and V. Trappe$^2$} 
\affiliation{ $^1$Department of
Mathematics, King's College, Strand,
London, WC2R 2LS, UK \\
$^2$Physics Department, Universit\'e de Fribourg, 1700 Fribourg,
Switzerland\\
$^3$
Rudolf Peierls Centre for Theoretical Physics, 
University of Oxford, 1 Keble Road, Oxford, OX1 3NP, UK \\
$^4$Laboratoire des Verres UMR 5587, Universit\'e Montpellier II
and CNRS, 34095 Montpellier, France \\
$^5$GDPC UMR 5581, Universit\'e Montpellier II and CNRS,
34095 Montpellier, France \\
$^6$School of Physics and Astronomy, University of Nottingham,
Nottingham, NG7  2RD, UK}
\date{\today}
\begin{abstract}
We show by means of experiments, theory and simulations, that the slow
dynamics of coarsening systems displays dynamic heterogeneity
similar to that observed in glass-forming systems.  
We measure dynamic heterogeneity via novel multi-point
functions which quantify the emergence of dynamic, as opposed to
static, correlations of fluctuations.
Experiments are performed on a coarsening foam using Time
Resolved Correlation, a recently introduced light scattering method. 
Theoretically we study the Ising model, and present exact results in one
dimension, and numerical results in two dimensions.
For all systems the same dynamic scaling of fluctuations with domain
size is observed.
\end{abstract}


\maketitle 

Glassy and jammed materials display similar phenomenology, characterized in
particular by slow and nonequilibrium dynamics, whose
microscopic origin is still being actively 
investigated~\cite{lezouches,debe,LiuNature1999}.
Recent research has shown that despite the absence of static ordering, 
glass-formers exhibit 
non-trivial spatial correlations of the local 
dynamics, resulting in
dynamic heterogeneity~\cite{dh5,dh1}.
Here, we take the view that slow dynamics is intrinsically associated to
dynamic heterogeneity, as suggested by studies of the glass 
transition~\cite{dh3}.

Dynamic heterogeneity has to be measured by means of 
statistical correlators that probe more than two points 
in space and time. An ideal experiment or
calculation would 
compare local configurations around position $r$ at times 
$t$ and $t+\dt$
via a two-time quantity, $F(r,t,\dt)$. 
Traditionally, only the dynamics averaged over
$t$, $r$, or thermal histories is discussed,
$\fav (\dt) = \langle \frac{1}{V} \int_V d^d r F(r,t,\dt) \rangle$.
In a glass-former, $\fav$ could be for example
the self-intermediate scattering function at a given 
wavevector.  
By contrast, our goal is to detect spatial correlations of the local
dynamics. A natural correlator is~\cite{dh5,dh1,dh3}
\begin{equation}
C(r, \Delta t) =
\frac{1}{V} \int_V d^d r' 
\langle F(r',t,\Delta t) F(r'+r,t,\Delta
t) \rangle - \fav^2(\Delta t), \label{cr}
\end{equation} 
built from two-point, two-time quantities. It is easier 
to measure the volume integral of (\ref{cr}), a dynamic
susceptibility $\chi(\dt) \equiv \int_V d^d r \, C(r, \dt)$, which
can be rewritten as the variance of the fluctuations of the
two-time dynamics,
\begin{equation}
\chi(\Delta t) = V \left[
\left\langle
\left( \frac{1}{V} \int_V d^d r F(r,t,\Delta t) \right)^2
\right\rangle
- \fav^2(\dt)
\right]
. \label{chi}
\end{equation}
Physically, dynamic fluctuations increase when the number of 
independent dynamic objects decreases, but normalizations 
ensure that $\chi$ remains finite in
the thermodynamic limit, except at a dynamic critical point, 
as discussed for supercooled liquids~\cite{dh2}.

\begin{figure}[b]
\psfig{file=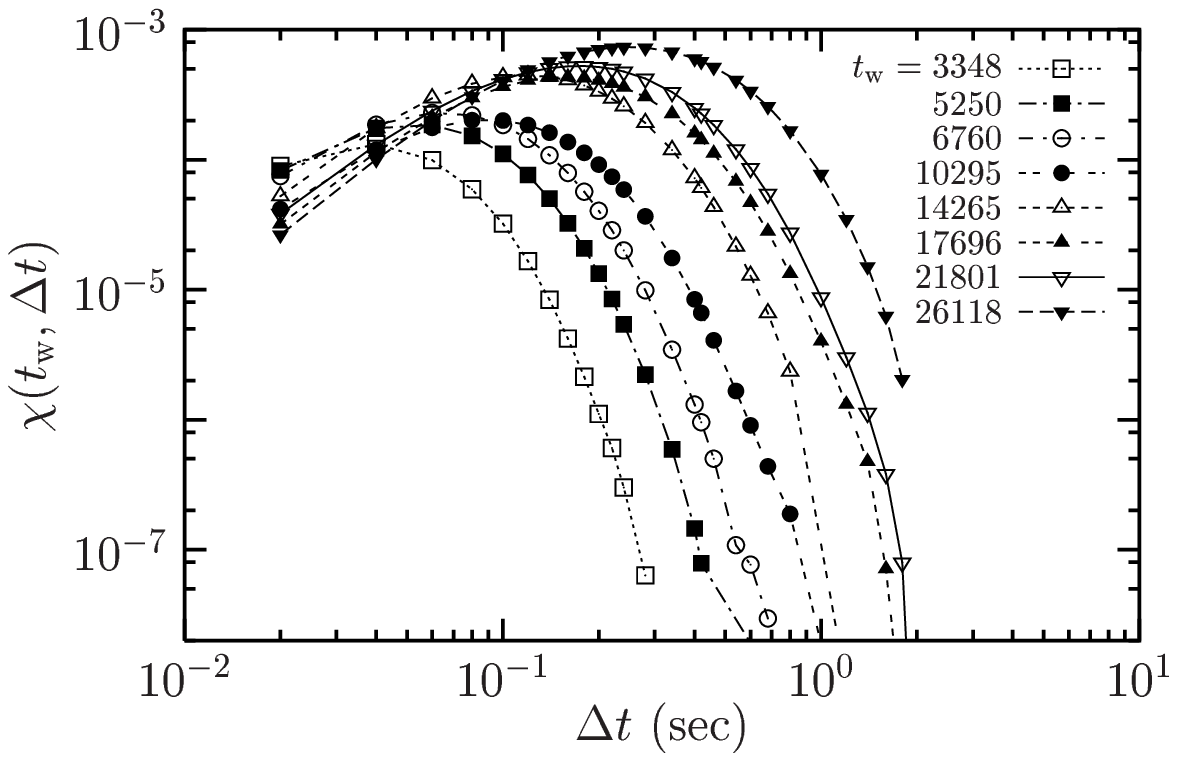,width=8.cm}
\psfig{file=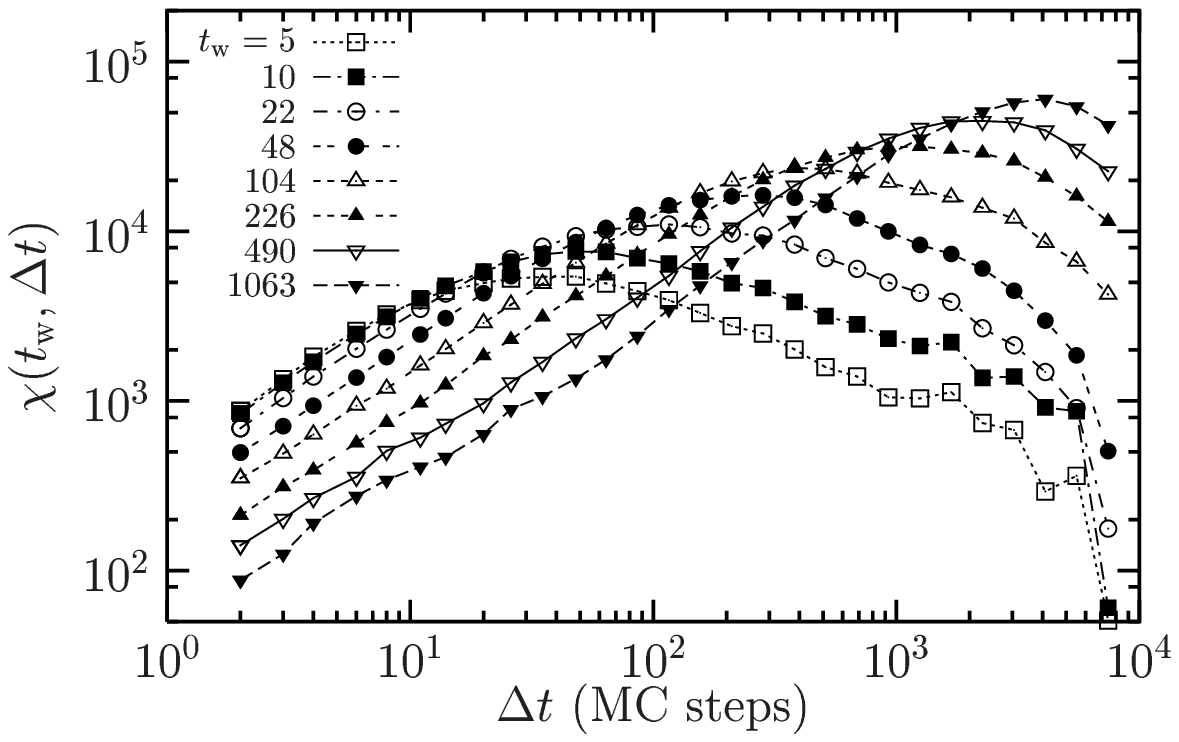,width=8.5cm}
\caption{\label{luca1} Dynamic
susceptibility $\chi(\tw,\dt)$ measured experimentally in a
coarsening foam (top), and numerically in the $d=2$ Ising model
(bottom) for various ages $\tw$.}
\end{figure}

In this paper, we focus on coarsening phenomena, as one of the
simplest and better understood physical situations characterized
by slow dynamics and aging~\cite{bray}.
We introduce new experimental and theoretical methods to access
$\chi$ and $C$, and show that 
coarsening systems display heterogeneous
dynamics similar to that observed at equilibrium 
in glass formers.
We perform experiments on a coarsening soft material, 
a dry foam, and we study the Ising model in one and two dimensions.
Although very different, these systems 
display the same dynamic scaling of fluctuations with domain size.
 
Experimentally, soft materials are well suited for investigating
dynamic heterogeneity because the relevant time and length scales
are much larger than in molecular systems, which greatly simplifies
detailed dynamic measurements. We use dynamic light scattering in
the strongly multiple scattering limit (Diffusing Wave
Spectroscopy, DWS \cite{DWSGeneral}) to probe the dynamics of a
shaving cream. The foam coarsens as the time $\tw$ since
its preparation increases. The time-averaged intensity autocorrelation
function measured by DWS, $g_2(\tw,\Delta t)-1$, decays
exponentially, with a characteristic
rate, $\Gamma(\tw)$, that decreases with $\tw$. This
results from spatially and temporally localized random
rearrangements~\cite{DurianScience1991}.

Traditional light scattering measurements use a point detector and
require an extended time average of the intensity correlation
function. Consequently, no information on the fluctuations of the
dynamics is accessible. To overcome this limitation, we use the
recently introduced Time Resolved Correlation
technique~\cite{LucaJPCM2003}. A charge-coupled device (CCD)
camera is used to record, at a constant rate, the speckle pattern
of the light scattered by the foam~\cite{Goodman}. The degree of
correlation, $c_I$, between speckles at times $\tw$ and
$\tw+\Delta t$ is calculated as
\begin{equation}
    c_I(\tw,\Delta t) = \frac{\langle I_p(\tw)I_p(\tw+\Delta
t)\rangle_p }
    {\langle I_p(\tw)\rangle_p \langle I_p(\tw+\Delta t)\rangle_p}-1~,
    \label{equ:cI}
\end{equation}
where $I_p(\tw)$ is the intensity measured at time $\tw$ by the
$p$-th CCD pixel and $\langle \cdot \cdot \cdot \rangle_p $
denotes an average over all pixels. Since different configurations
yield different speckle patterns, $c_I(\tw,\Delta t)$ quantifies
the degree of correlation of the foam between times $\tw$ and
$\tw+ \Delta t$. Because the experiments are performed in the
transmission geometry and in the strong multiple scattering
regime, each CCD pixel collects light coming from the whole
scattering volume. Therefore, $c_I(\tw,\Delta t)$ provides
time-resolved but spatially-integrated dynamical information. The
standard intensity correlation function is obtained via a further
average, $g_2(\tw,\Delta t)-1 = \langle c_I(\tw,\Delta t)
\rangle_{T}$, 
which is analogous to the correlator $\fav$ defined above.
Here, $\langle \cdot \cdot \cdot \rangle_{T}$
denotes an average over $[ \tw, \tw+T ]$, with $T \gg \Gamma^{-1}$, 
but short enough to prevent any significant
change of the dynamics due to coarsening. 
In pratice, we took $T < 0.05 \tw$ for all ages, 
and checked that our results do not depend on this choice.
Dynamic fluctuations are quantified via the variance of $c_I$, similarly to
Eq.~(\ref{chi}): $\chi(\tw, \Delta t) = \langle c_{I}^2(\tw,\Delta
t) \rangle_{T} - \langle c_{I}(\tw,\Delta t) \rangle_{T}^2$.

The dynamic susceptibility $\chi(\tw, \Delta t)$ measured during
the aging of the foam, from $\tw = 3350$ to $26200$ sec is shown
in Fig.~\ref{luca1}. Trivial contributions to the fluctuations due
to the CCD noise and the finite number of speckles have been
subtracted from the data~\cite{DuriSPIE2004}. For all $\tw$, we 
find that
$\chi$ exhibits a peak at time lags $\dt^\star(\tw)$ close to
$\Gamma^{-1}(t_{\rm w})$. Moreover, as the foam ages and coarsens, the
height of the peak, $\chi^\star(\tw) = \chi(\tw,\dt^\star)$
increases and its position shifts to larger times, 
in striking analogy with numerical observations in supercooled
liquids~\cite{dh5,dh1}. 
To our knowledge, no experimental measurement of the dynamic
susceptibility was so far reported.

A similar behavior can be observed in a very different
coarsening system. We study the dynamics of the 
Ising model on a regular lattice quenched
from a random state to the ferromagnetic phase. Domains of
positive and negative magnetization develop and grow with $\tw$,
in analogy with the bubbles of the foam. The Hamiltonian is $H = -
\sum_{\langle i,j \rangle} \s_i \s_j$, the sum being over nearest
neighbor pairs. From (\ref{cr}), one would naively study
$C_0(l-k,\tw,\dt)  =
     \langle \s_k(t) \s_k(\tw) \s_l(t) \s_l(\tw) \rangle -
     \langle \s_k(t) \s_k(\tw) \rangle^2$,
where $t=\dt+\tw$. 
However, $C_0$ is trivially dominated by equal-time
two-point correlations:
consider for instance the large time
limit, where $C_0(l-k,\tw,\dt\to\infty) \to \langle \sigma_k(t)
\sigma_l(t) \rangle \langle \sigma_k(\tw) \sigma_l(\tw)\rangle$.
The appropriate correlator to consider is instead~\cite{tobe}
\begin{eqnarray}
   C(l-k,\tw,\dt) & = & -
     \langle \s_k(t) \s_k(\tw) \s_l(t) \s_l(\tw) \rangle
     \nonumber \\
& & + \langle \s_k(t) \s_k(\tw) \rangle \langle \s_l(t) \s_l(\tw)
\rangle \nonumber \\
   & & +
     \langle \s_k(t) \s_l(t) \rangle \langle \s_k(\tw) \s_l(\tw) \rangle
     \nonumber \\
   & & -
     \langle \s_k(t) \s_l(\tw) \rangle \langle \s_k(\tw) \s_l(t) \rangle,
     \label{equ:Cndef}
\end{eqnarray}
where the relative signs are consequences of the fermionic
nature of fluctuations in the Ising model.
The corresponding susceptibility is defined as in Eq.~(\ref{chi}),
\begin{equation}
   \chi(\tw,\dt)=\sum\limits_n C(n,\tw,\dt).
   \label{equ:Cdef}
\end{equation}
Note that (\ref{equ:Cdef}) is not positive definite, but to 
ease the comparison with previous research, 
definitions (\ref{equ:Cndef}) and (\ref{equ:Cdef}) are chosen so that these
quantities are eventually positive.

Dynamic fluctuations in the $d=2$ Ising model are measured in
Monte Carlo (MC) simulations.
The simulated system, $L$, must be
large enough that the mean domain size, $R(\tw)$, satisfies
$R(\tw) \ll L$ at all times, but small enough not to average out
the fluctuations. We used $L=600$ and averaged the results
over $2 \cdot 10^3$ independent initial conditions, the total
simulated time being $10^4$ MC steps. 
Due to computational limitations, we could only
measure $\chi(\tw,\dt)$ as the linear integral 
of (\ref{equ:Cndef}) measured along the 
$x$ and $y$ axes.
As seen in Fig.~\ref{luca1}, the observed dynamical
fluctuations are similar to those observed for the foam.

This similarity lies in the common physical mechanism
responsible for the slow, heterogeneous dynamics: the domain growth
driven by the reduction of interfacial energy.
As the system coarsens, the size of the regions that
undergo correlated rearrangements increases, so that 
the number of independent regions in the probed volume, $N(\tw)$, 
decreases, which in turn increases the amplitude of the dynamic fluctuations. 
One would therefore expect that $\chi^\star(\tw) \sim N^{-1}(\tw) 
\sim R^d(\tw)$.

To substantiate this prediction, we first
consider an analytically solvable case. 
We study the non-equilibrium dynamics of the $d=1$
Ising model. The spin chain evolves according to standard Glauber 
rules~\cite{Glauber63}.
From general expressions for two-time, multispin
correlation functions following a quench derived
in~\cite{MaySol04}, we compute exactly the quantities
(\ref{equ:Cndef}) and (\ref{equ:Cdef}). Although straightforward
in principle, actual calculations require technically involved
algebraic manipulations, detailed in~\cite{tobe}.

Consider first equilibrium dynamics which we expect to
be dynamically homogeneous. Indeed, we find that the
correlation (\ref{equ:Cndef}) vanishes exactly,
$C^{\eq}(n,\dt)=\lim_{\tw \to \infty} C(n,\tw,\dt) = 0$, for
arbitrary distances $n$, times $\dt$, and temperatures $T>0$. 
This implies that two-time multispin correlations 
factorize into two-spin static correlations, 
and shows that the correlator (\ref{equ:Cndef}) is well suited 
to revealing the existence of non-trivial, dynamic, 
out of equilibrium correlations.

\begin{figure}
   \epsfig{file=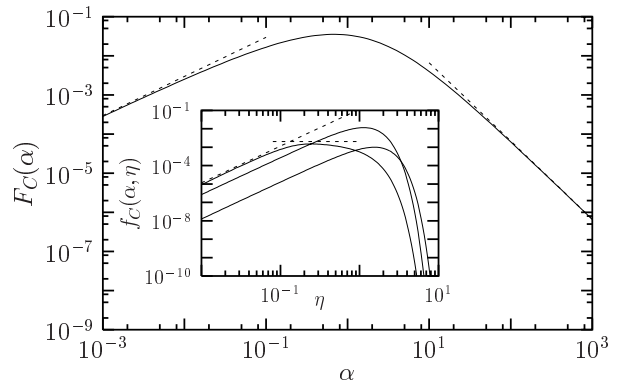,width=8cm}
   \caption{\label{fig:CorSum} Exact scaling function
     of Eq.~(\ref{equ:Cagingscaling}) for the dynamic 
susceptibility of the Ising chain. Dashed lines show the asymptotic
behavior, $F_C(\alpha \to 0)
\sim \alpha$ and $F_C(\alpha \to \infty) \sim \alpha^{-2}$.
Inset: Exact scaling function of Eq.~(\ref{scaleinvariance}), for
$\alpha=10^{-2}$, 1 and 10 (from top to bottom on the left). 
Dashed lines represent the asymptotic behavior estimated via random 
walk arguments.}
\end{figure}

The situation is more interesting in
the coarsening regime. We focus on the scaling
behavior for large times $\dt,\tw \to \infty$ and distances $n \to
\infty$, with scaling variables $\alpha=\dt/\tw$ and $\eta= n
/R(\tw)$ fixed, where $R(\tw) \sim \sqrt{\tw}$. 
One finds scale invariance,
\begin{equation}
   C(n,\tw,\dt) \sim
f_C\left(\frac{\dt}{\tw},\frac{n}{R(\tw)}\right),
\label{scaleinvariance}
\end{equation}
where the scaling function
$f_C(\alpha,\eta)$ is given in~\cite{tobe}.
From (\ref{scaleinvariance}),
the susceptibility (\ref{equ:Cdef}) scales as
\begin{equation}
   \chi(\tw,\dt) \sim R(\tw) \, F_C\left( \frac{\dt}{\tw} \right),
\label{equ:Cagingscaling}
\end{equation}
with $F_C(\alpha) = \int \dd\eta \, f_C(\alpha,\eta)$.
Both scaling functions are displayed in Fig.~\ref{fig:CorSum}.
In particular, $\chi(\tw,\dt)$ qualitatively resembles 
the results in Fig.~\ref{luca1}, and
Eq.~(\ref{equ:Cagingscaling}) shows that the
peak height scales as expected, $\chi^\star(\tw) \sim R(\tw)$.

To develop a more direct understanding of (\ref{scaleinvariance}),
we note that the dynamics of the Ising spin chain can be mapped to
a diffusion-limited annihilation process, where one studies the
dynamics of the walls separating domains of opposite 
magnetization rather than that of 
the spins themselves~\cite{Santos97}. 
Trajectories of the walls are random walks that
annihilate whenever they meet.
In a space-time diagram~\cite{dh3}, the spins 
$\sigma_k(\tw)$, $\sigma_l(\tw)$, $\sigma_k(t)$ and
$\sigma_l(t)$ occupy the corners of a rectangle of size $n = k
- l$ by $\dt = t - \tw$, and the spin products in
(\ref{equ:Cndef}) are determined by the parity of the number of
random walks crossing the relevant edge. Labeling edges in
the order left - right - top - bottom we denote, for example, an
odd number of random walkers crossing the left and bottom edges 
as 1001. Because walls
annihilate in pairs, the number of walkers crossing the
rectangle is even, so that there are only 8 possible situations.
In terms of the corresponding probabilities, (\ref{equ:Cndef}) may
be reexpressed as $C(n,\tw,\dt) = 8 ( p_{0101}\,p_{1010} +
p_{0110}\,p_{1001} - p_{0000}\,p_{1111} - p_{0011}\,p_{1100})$.
These probabilities can be evaluated to leading order via standard
random walk arguments~\cite{tobe}, when the number of trajectories
crossing the rectangle is small, i.e. for diluted walls, $n \ll R(\tw)$,
and short time delays, $\dt \ll \tw$.
For $\dt \ll n^2 \ll \tw$, $C(n,\tw,\dt)$ converges to an
$n$-independent plateau of height $(2/\pi^2) \alpha$, 
see Fig.~\ref{fig:CorSum}.
When $n^2 \ll \dt \ll \tw$, on the other hand,
$C(n,\tw,\dt) \approx \pi^{-1}(1-2/\pi) n^2/\tw$, implying that
$C$ grows like $\eta^2$ with a $\dt$-independent amplitude, see
Fig.~\ref{fig:CorSum}.
The random walk picture becomes too complicated when either $n^2$ or
$\dt$ are large compared to $\tw$ and we refer to our exact
results in this regime~\cite{tobe}. For $n^2\ll\tw\ll\dt$, the
$\eta^2$-dependence found above for $n^2\ll\dt\ll\tw$ persists,
but now with an amplitude that decreases as $\alpha^{-2}$. For
large $n^2 \gg \tw$, finally, we find a Gaussian cutoff in $n$
with a width  of order $\sqrt{\tw}$ for both $\dt \ll \tw$ and
$\dt \gg \tw$.

The asymptotic behavior of $\chi(\tw,\dt)$ follows
from the above discussion.
For $\dt\ll\tw$, the integral is dominated by the plateau region
of $f_C$. Since the plateau grows
as $\alpha$, so does (\ref{equ:Cagingscaling}). For $\dt\gg\tw$,
on the other hand, the decrease in the amplitude of the
$\eta^2$-part of $C$ controls the variation of $\chi$ which decays
therefore as $\alpha^{-2}$. Rigorous analysis shows indeed
$F_C(\alpha \to 0) \sim 4\left(\sqrt{2}-1\right)\pi^{-3/2} \alpha$
and $F_C(\alpha \to \infty) \sim (8/5)\left(8\sqrt{2}-9\right)
\pi^{-3/2} \alpha^{-2}$, see Fig.~\ref{fig:CorSum}. 

\begin{figure}
   \epsfig{file=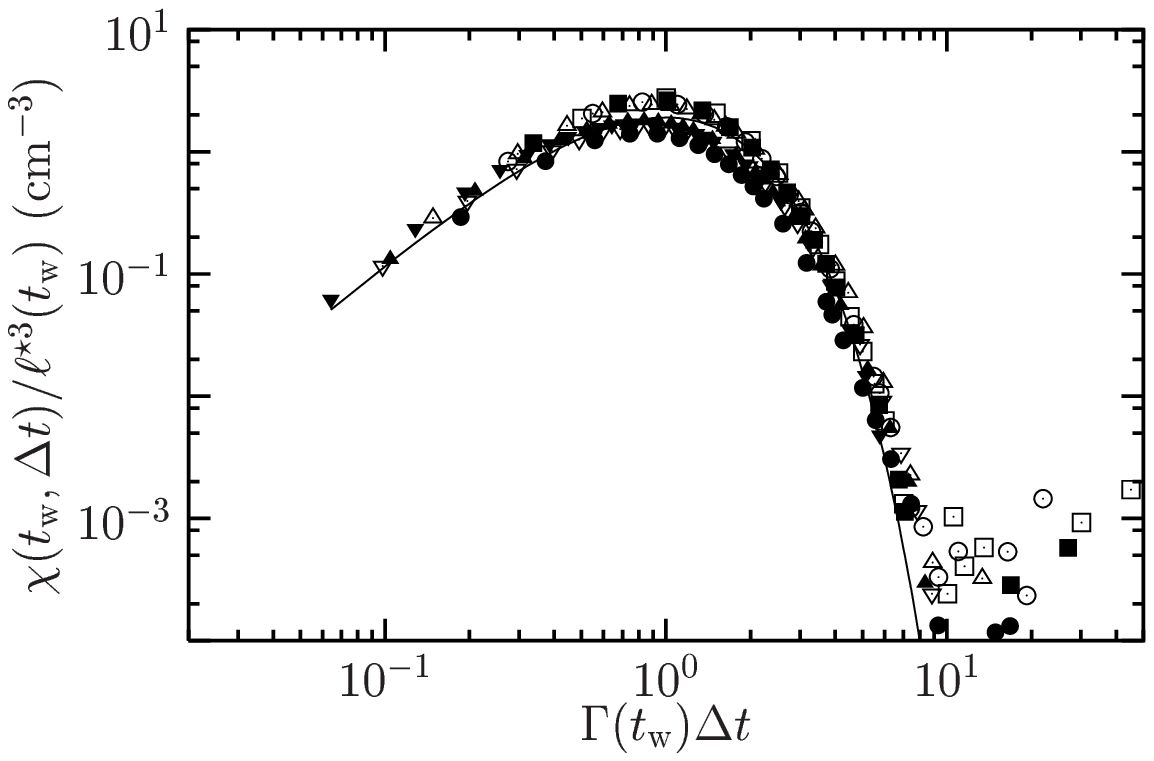,width=8.5cm}
   \epsfig{file=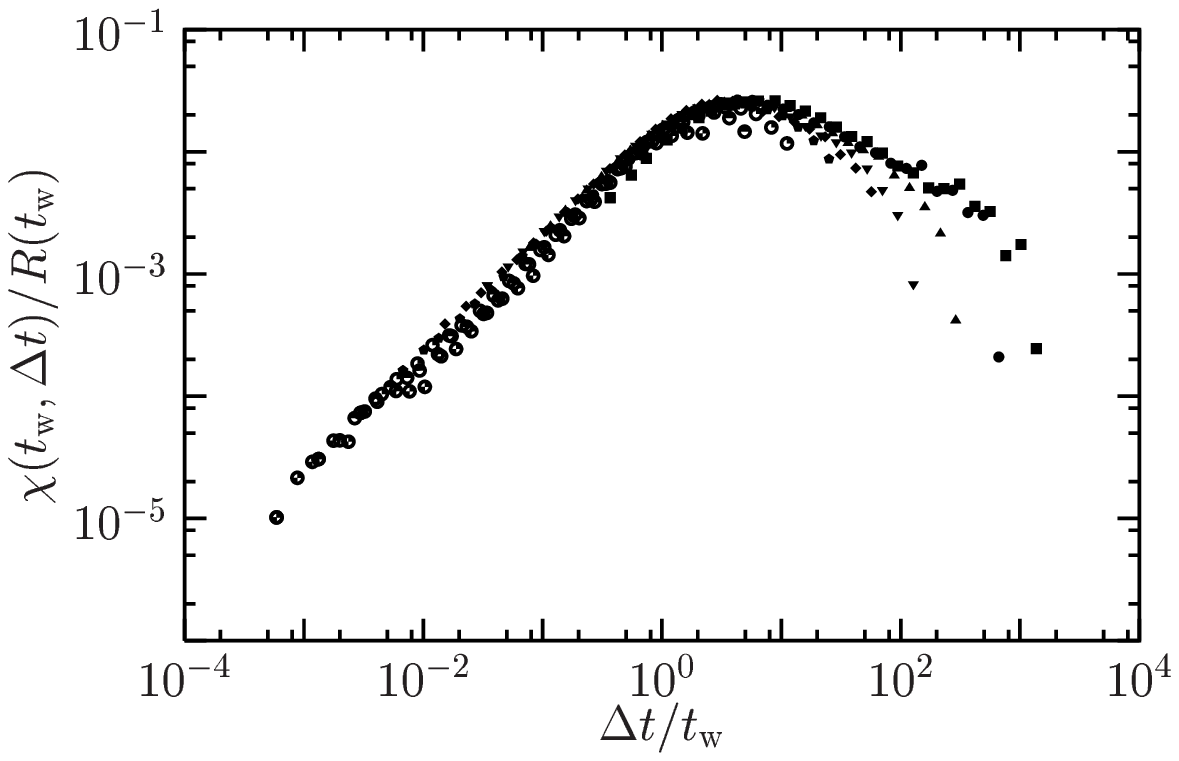,width=8.5cm}
   \caption{\label{scaling} Dynamic
scaling of all dynamic susceptibilities in Fig.~\ref{luca1} (same
symbols). Top: experiments on a foam, the line is the scaling function of
Eq.~(\ref{equ:varcI}).
Bottom: numerical simulations of the $d=2$ Ising model.
}
\end{figure}

For the $d=2$ Ising model and the foam no analytical results 
are available. Instead, we test the scaling of $\chi(\tw,\dt)$
by measuring $R(\tw)$ and the characteristic
relaxation time. In simulations, $R(\tw)$
is determined from the decay at large distance of the equal-time structure
factor, while the characteristic decay rate scales as $1/\tw$, 
$\Gamma(\tw) \sim \tw^{-1}$~\cite{bray}.
In the simulations,
$\chi$ is a linear integral over distance, 
as mentioned above. Thus,
one expects $\chi^\star(\tw) \sim R(\tw)$. 
For the foam, $R(\tw)$ scales as the bubble size, which is also
proportional to the photon transport mean free path,
$\ell^\star(\tw)$~\cite{DurianScience1991}, which we obtain
independently from transmission measurements~\cite{DWSGeneral}.
We expect therefore $\chi^\star(\tw) \sim {\ell^\star}^3 (\tw)$. 

As seen in Fig.~\ref{scaling}, not only does the peak of 
the dynamic susceptibility scale as expected, but data for all times
collapse onto a master curve, a scaling behavior analogous 
to that found for the Ising chain, Eq.~(\ref{equ:Cagingscaling}).
For the foam, the shape of the master curve can be explained 
by noting that at all times, $c_I(\tw,\Delta t) \sim
\exp [ - \gamma(\tw) \dt]$, where $\gamma(\tw)$ is a decay rate
fluctuating with the number of rearrangements, of mean
$\Gamma(\tw) = \langle \gamma(\tw) \rangle_{T}$. To leading
order in the variance of $\gamma(\tw)$, $\sigma^2(\tw) \equiv
\langle \gamma(\tw)^2 \rangle_{T} - \langle \gamma(\tw)
\rangle_{T}^2 \sim \ell^{*3}(\tw) \Gamma^2(\tw)$, one can
estimate
\begin{equation}
    \chi(\tw,\dt) \sim \ell^{*3}(\tw)
    \big[ \Gamma(\tw) \dt \big]^2
    e^{-2  \Gamma(\tw) \dt}.
    \label{equ:varcI}
\end{equation}
The solid line in the top panel of Fig.~\ref{scaling} shows
that the scaling function $f(x) = x^2 \exp(-2x)$, suggested
by~(\ref{equ:varcI}), is in very good agreement with the
experimental data.

In summary, we have defined and analyzed dynamic spatial correlators and
susceptibilities revealing the growth with time of 
dynamic heterogeneity in coarsening systems. The increasing dynamical
fluctuations are consequences of a reduced number of
independent dynamical domains ---bubbles or magnetized domains---
as coarsening proceeds.
The non-fractal morphology of the
domains implies a particularly simple scaling of spatial
correlators, and therefore of their volume integral,  
$\chi(\tw,\dt) \sim R^d(\tw) f(\Gamma(\tw) \dt)$. 
However, dynamic heterogeneity is also found
in systems where dynamic domains are believed to have a
fractal morphology~\cite{dh5,dh1}.
In this case, a more complicated dynamic scaling can
be expected~\cite{dh2,dh5}.
The techniques presented in this work
will help identify and characterize dynamic heterogeneity, an
endeavor that appears as a crucial step in gaining a better
understanding of the dynamic slowing down of many glassy and
jammed materials.

\begin{acknowledgments}
We thank P. Ballesta and A. Lef\`evre for useful discussions. We
acknowledge financial and numerical support from 
EPSRC grants 00800822, GR/R83712/01 and GR/S54074/01, 
E.U. grant No.\ HPMF-CT-2002-01927, 
SNF Grant 2100-066920,
CNRS (PICS 2410), 
French Minist\`ere de la Recherche (ACI Jeunes Chercheurs), 
ESF Program SPHINX,
Nuffield Grant NAL/00361/G, 
\"{O}sterreichische Akademie der Wissenschaften,
Worcester College Oxford, 
and Oxford Supercomputing Center.
\end{acknowledgments}

\end{document}